\documentclass[12pt]{article}
\pdfoutput=1
\usepackage{jhep-mod}
\usepackage{bm}
\usepackage{amssymb}% http://ctan.org/pkg/amssymb
\usepackage{pifont}% http://ctan.org/pkg/pifont
\usepackage{slashed} % for Dirac operator
\usepackage{slantsc} % slanted smallcaps...
\begin{document}
%-------------------------------------------------------------------------------------------
\title{Conformally Friedmann--Lema\^\i{}tre--Robertson--Walker cosmologies}
%-------------------------------------------------------------------------------------------
\author{Matt Visser}
\affiliation{School of Mathematics, Statistics, and Operations Research, \\
Victoria University of Wellington, PO Box 600, Wellington 6140, New Zealand}
\emailAdd{matt.visser@msor.vuw.ac.nz}
%-------------------------------------------------------------------------------------------
\abstract{
In a universe where, according to the standard cosmological models, some 97\% of the total mass-energy is still ``missing in action'' it behooves us to spend at least a little effort critically assessing and exploring radical alternatives. Among possible, (dare we say plausible),  nonstandard but superficially viable models, those
spacetimes conformal to the standard Friedmann--Lema\^\i{}tre--Robertson--Walker class of cosmological models play a very special role --- these models have the unique and important property of permitting large non-perturbative geometric deviations from Friedmann--Lema\^\i{}tre--Robertson--Walker cosmology without unacceptably distorting the cosmic microwave background. 
Performing a ``cosmographic'' analysis, (that is, temporarily setting aside the Einstein equations, since the question of whether or not the Einstein equations are valid on galactic and cosmological scales is essentially the same question as whether or not dark matter/dark energy actually exist), and using both supernova data and information about galactic structure, one can nevertheless place some quite significant observational constraints on any possible conformal mode --- however there is still an extremely rich range of phenomenological  possibilities for both cosmologists and astrophysicists to explore.

\bigskip
\noindent
10 February 2015; 17 February 2015; 21 April 2015; \LaTeX-ed \today;\\
arXiv:1502.02758 [gr-qc]. 
}
%-------------------------------------------------------------------------------------------
%\pacs{98.80.-k  98.80.Jk  95.36.+x 95.30.Sf }
%-------------------------------------------------------------------------------------------
%98.80.-k Cosmology
%98.80.Jk Mathematical and relativistic aspects of cosmology
%95.36.+x Dark energy 
%95.30.Sf Relativity and gravitation
%-------------------------------------------------------------------------------------------
\maketitle
%-------------------------------------------------------------------------------------------
\def\d{{\mathrm{d}}}
%-------------------------------------------------------------------------------------------
\section{Background} 
%-------------------------------------------------------------------------------------------
\parskip5pt
Looking out into the universe, one notes that the cosmic microwave background [CMB] is extremely smooth, while in marked contrast the recent distribution of matter is extremely (non-perturbatively) clumpy. This raises the natural question: Does the large-scale geometry of spacetime more closely track the CMB, or does the geometry more closely track the distribution of matter? Is the cosmological spacetime geometry smooth? Or is it (non-perturbatively) clumpy?

For the last eighty-odd years, cosmological spacetime geometry has, in the main, been viewed as the study of Friedmann--Lema\^\i{}tre--Robertson--Walker [FLRW] spacetimes, with \emph{small} perturbations in the geometry~\cite{Liddle, Weinberg1, Weinberg2}. From the standard perspective, keeping the geometry smooth is normally deemed essential to keeping the CMB smooth. Observationally, (excluding the dipole term), the CMB is smooth to 1 part in $10^5$, and the traditional view is that this implies that the spacetime geometry should be smooth to a similar level~\cite{WMAP1, WMAP3, WMAP5, WMAP7, WMAP9, WMAP9b,  Bean:2003, Wang:2006}. In contrast, the distribution of visible matter in the ``recent'' ($z \lesssim 5$) universe is extremely (non-perturbatively) clumpy; similarly the distribution of dark matter, (but not necessarily dark energy), is known to be extremely (non-perturbatively) clumpy~\cite{Liddle, Weinberg1, Weinberg2}.  This (non-perturbative) clumpiness of the matter distribution is so extreme that there have been repeated suggestions in the literature that the deviations from FLRW \emph{geometry} should also in some sense be non-perturbative; but suggestions along these lines are difficult to then render compatible with the observed smoothness of the CMB~\cite{WMAP1, WMAP3, WMAP5, WMAP7, WMAP9, WMAP9b,  Bean:2003, Wang:2006}.

There is one way, \emph{and only one way}, in which the (relatively recent, $z \lesssim$~1000) \emph{geometry} can depart significantly (non-perturbatively) from FLRW spacetime without significantly (and disastrously) impacting the smoothness of the CMB, and that is to consider (non-pertubative) conformal deformations of FLRW cosmology. Conformal deformations of spacetime geometry are unique in that they do not affect null geodesics (photon paths), so that one can introduce an arbitrary amount of conformal warping and stretching between the here-and-now and the surface of last scattering ($z \approx$ 1100) without in any way affecting the angular distribution and correlations of the photons in the CMB. Because the affine parameter of null geodesics is not invariant under conformal deformations, there will be an effect on the overall temperature of the CMB, but we shall soon see that this will be an overall temperature shift, not an angular distortion.  

In counterpoint, there has also been a recent resurgence of interest in cosmography, (\emph{aka} cosmokinetics), where, (in view of the very significant uncertainties in our current understanding of cosmology), one delays dynamical considerations as long as possible, by extracting as much information as possible from the interplay between observation and the purely kinematic aspects of the (perturbed) FLRW geometry~\cite{Blandford:2004, Visser:2004, Cattoen:2007-1, Cattoen:2007-2, Cattoen:2008, Visser:2009, Vitagliano:2009, Xia:2011, Capozziello:2008, Guimaraes:2009, Sathyaprakash:2009, Guimaraes:2010, Izzo:2010, Capozziello:2011, Lavaux:2011, Bamba:2012}.  Adopting the cosmographic framework, and in particular working with both the supernova data~\cite{SS:1998-1, SS:1998-2, SS:1998-3, Riess:2000, Tonry:2003, WoodVasey:2007, Davis:2007, Kowalski:2008, Rubin:2008}, and basic features of galactic dynamics, we shall soon see that it is possible to place significant but not overwhelming constraints on any possible non-perturbative conformal deformation from the standard FLRW cosmology. 

The conformally FLRW [CFLRW] spacetimes considered in this article are somewhat loosely related to the so-called ``swiss cheese'' cosmologies~\cite{Biswas:2007, Marra:2007-1, Marra:2007-2, Vanderveld:2008, Valkenburg:2009, Flanagan:2011}, and are also natural modifications of Wiltshire's ``timescape'' cosmology~\cite{Wiltshire:2007-1, Wiltshire:2007-2, Wiltshire:2007-3, Leith:2007, Wiltshire:2008, Wiltshire:2009, Smale:2010, Wiltshire:2011-1, Smale:2011, Wiltshire:2011-2, Duley:2013, Wiltshire:2013, Wiltshire:2014}. They are also quite naturally related to the Lema\^\i{}tre--Tolman--Bondi [LTB] cosmologies~\cite{Barrow:1984, Rasanen:2004, Chung:2006, Paranjape:2006, Vanderveld:2006, Alexander:2007, GarciaBellido:2008, Yoo:2008, Enqvist:2007, Romano:2009, Krasinski}, and so serve to provide a single unifying framework against which all of these non-standard cosmological models can be contrasted. 

%-------------------------------------------------------------------------------------------
\section{CFLRW cosmography} 
%-------------------------------------------------------------------------------------------
Standard FLRW cosmological models consider a spacetime geometry defined by the line element:
\begin{eqnarray}
\d s^2 &=& - \d t^2 + a(t)^2 \left\{ {\d r^2\over1-kr^2} + r^2\left(\d\theta^2 + \sin^2\theta \; \d\phi^2\right)\right\} 
+ \hbox{``small perturbations''}.\qquad
\end{eqnarray}
Consider instead the conformally distorted CFLRW line element:
\begin{eqnarray}
\d s^2 &=& \exp\{2\Theta(t,r,\theta,\phi)\} 
\left[ - \d t^2 + a(t)^2 \left\{ {\d r^2\over1-kr^2} + r^2\left(\d\theta^2 + \sin^2\theta \; \d\phi^2\right)\right\}  \right] 
\nonumber\\
&&
+ \hbox{``small perturbations''}.
\end{eqnarray}
Here the conformal function $\Theta(t,\vec x)$ is permitted to be non-perturbatively large, and is permitted to depend (arbitrarily) on both space and time. 
To unambiguously separate out the local physics that is encoded in $\Theta(t,\vec x)$ from the overall expansion of the universe encoded  in $a(t)$, it is useful to impose a universe-wide spatial average condition such as (for instance) $\langle e^\Theta \rangle_\mathrm{spatial}=1$. 

Other possible ``gauge choices'' could be mooted. As an example, one could pick a specific timelike curve $(t,\vec x_0(t))$; representing for instance our own galaxy's past history, and arrange the interplay between $\Theta(t,\vec x) $ and $a(t)$ to set $\Theta(t,\vec x_0(t))=0$. This amounts to picking ``proper time gauge'' for one specific observer. The discussion below will be set up in such a manner as to, (as much as possible),  be independent of such gauge choices.
To explicitly see the need for such a ``gauge choice'' one could for instance go to conformal time $\eta$ by defining $\d\eta = \d t/a(t)$ in which case the CFLRW line element can be written as
\begin{eqnarray}
\d s^2 &=& \exp\{2\Theta(\eta,r,\theta,\phi)\} a(\eta)^2 
\left[ - \d \eta^2 + \left\{ {\d r^2\over1-kr^2} + r^2\left(\d\theta^2 + \sin^2\theta \; \d\phi^2\right)\right\}  \right] 
\nonumber\\
&&
+ \hbox{``small perturbations''}.
\end{eqnarray}
By now defining $\Phi = \Theta + \ln a$ we can completely absorb the ``size of the universe'' into the conformal mode and rewrite the CFLRW line element as
\begin{eqnarray}
\d s^2 &=& \exp\{2\Phi(\eta,r,\theta,\phi)\} 
\left[ - \d \eta^2 +  \left\{ {\d r^2\over1-kr^2} + r^2\left(\d\theta^2 + \sin^2\theta \; \d\phi^2\right)\right\}  \right] 
\nonumber\\
&&
+ \hbox{``small perturbations''}.
\end{eqnarray}
While this form of the CFLRW line element is very convenient for purely theoretical purposes, observational comparisons are more easily facilitated by splitting $\Phi(\eta,r,\theta,\phi)$ into an ``overall size'' $a(\eta)$ component and  a ``local stretching'' mode $\Theta(\eta,r,\theta,\phi)$; the universe-wide spatial average  $\langle e^\Theta \rangle_\mathrm{spatial}=1$ is one way of doing so. The ``proper time gauge'' is another way of doing so. 

Note furthermore that the $k=0$ unperturbed CFLRW spacetime (setting the ``small perturbations'' to zero)  is manifestly conformally flat. Even for the $k=\pm1$ unperturbed CFLRW spacetimes the Weyl tensor is identically zero, so coordinates can be chosen to  make these spacetimes manifestly locally conformally flat. Unfortunately those coordinates mix space and time in a messy way, so while possibly useful for theoretical purposes, such coordinates are awkward for observational comparisons.

Temporarily ignoring the ``small perturbations'',  CFLRW spacetime has exactly the same null geodesics as ordinary FLRW spacetime. That is, photon propagation in this CFLRW spacetime is completely equivalent to that in ordinary FLRW spacetime, thus in principle permitting arbitrarily large (non-perturbative) deviations from FLRW cosmology without disturbing the CMB. 
More precisely, one would  want $\Theta(t,\vec x)$ to be ``perturbatively small'' at the surface of last scattering, (so that $\Delta [\Theta_\mathrm{last\,scattering}]\lesssim 10^{-5}$, at redshift $z\approx 1100$), but one can then permit $\Theta(t,\vec x)$ to become non-perturbatively large in the more recent universe ($z\ll1100$) without distorting the spectral shape and angular distribution of the CMB. 
In these CFLRW spacetimes the Weyl tensor should remain perturbatively small (compatible with observed gravitational lensing) all the way back to last scattering.

%-------------------------------------------------------------------------------------------
\section{Other cosmographies} 
%-------------------------------------------------------------------------------------------

Let us now contrast the CFLRW cosmography against some other cosmographies based on the LTB, Swiss-cheese, and timescape cosmologies. 

\enlargethispage{40pt}
The Lema\^\i{}tre--Tolman--Bondi [LTB] metric can be written in many ways. One form particularly useful from a cosmographic point of view is this
\begin{equation}
\d s^2 = - \d t^2 + A(r,t)^2 \d r^2 + B(r,t)^2\left(\d\theta^2 + \sin^2\theta \; \d\phi^2\right).
\end{equation}
This is simply the most general spherically symmetric time-dependent spacetime written in ``comoving coordinates", such that the timelike curves generated by holding $(r,\theta,\phi)$ fixed and letting $t$ vary are geodesics. In LTB cosmology (as opposed to LTB cosmography) one immediately chooses a matter model (typically comoving dust) and imposes the Einstein equations; for current purposes, I shall eschew this second step. (The history of these LTB models [\emph{aka} L--T models] is quite complex; see Krasi\'nski for historical details~\cite{Krasinski}.) The LTB cosmographies generically have non-zero Weyl tensor; only for the FLRW sub-class  (eg: $A(r,t)\to a(t)/\sqrt{1-k r^2}$, and $B(r,t)\to r a(t)$), does the Weyl tensor vanish.  Because of this feature the CMB photons will be scattered (by the Weyl tensor) and the CMB will be distorted. This is why in LTB models our galaxy is typically placed near the centre of the universe, to keep the distortions in the CMB  (and the Hubble flow) somewhat under control. 

The Swiss-cheese cosmographies are constructed by inserting multiple non-overlapping spherically symmetric inclusions into some FLRW background. Three types of spherically symmetric inclusion are common: 
\begin{itemize}
\item Schwarzschild vacuole.
\item LTB inclusion.
\item FLRW inclusion with $k$ distinct from that of the background.
\end{itemize}
The Schwarzschild and LTB insertions have non-zero Weyl tensor, and so certainly scatter the CMB. More subtly a FLRW insertion into a  FLRW background with different $k$ will have a delta-function supported Weyl tensor at the transition zone from one FLRW to another.  (Similarly there will be a delta-function contribution to the Weyl tensor at the transition zone for Schwarzschild and LTB insertions.)
In all three cases CMB photons will be scattered (by the Weyl tensor) and the CMB will be distorted. This is why in Swiss-cheese models our galaxy is typically placed near the centre of one of the inclusions, to keep the distortions in the CMB  (and the Hubble flow) somewhat under control. 

The timescape cosmographies are more subtle~\cite{Wiltshire:2007-1, Wiltshire:2007-2, Wiltshire:2007-3, Leith:2007, Wiltshire:2008, Wiltshire:2009, Smale:2010, Wiltshire:2011-1, Smale:2011, Wiltshire:2011-2, Duley:2013, Wiltshire:2013} --- in those models the metric is taken to be resolution-dependent. (That is, dependent on the distance scale on which one averages.) This has some resemblance to the particle physics concept of a resolution-dependent ``running coupling constant''. 
\begin{itemize}
\item When averaging over a void region, with a resolution smaller than that of the void, one takes the running metric to be $k=-1$ FLRW, for convenience in the form
\begin{equation}
\d s^2 = - \d t_\mathrm{void}^2 + a(t_\mathrm{void})^2 \left[ \d r_\mathrm{void}^2 + \sinh^2(r_\mathrm{void})\left(\d\theta^2 + \sin^2\theta \; \d\phi^2\right) \right].
\label{E:void}
\end{equation}

\item When averaging over a wall/filament/knot region, with a resolution smaller than that of the wall/filament/knot, one takes the running metric to be $k=0$ FLRW.
\begin{equation}
\d s^2 = - \d t_\mathrm{wall}^2 + a(t_\mathrm{wall})^2 \left[ \d r_\mathrm{wall}^2 + r_\mathrm{wall}^2\left(\d\theta^2 + \sin^2\theta \; \d\phi^2\right) \right].
\label{E:wall}
\end{equation}

\item When averaging over a horizon-scale region that contains many walls/filaments/knots and voids one takes the running metric to be of the form
\begin{equation}
\d s^2 = - \d t^2 + A(t)^2 \d r^2 + B(r,t)^2\left(\d\theta^2 + \sin^2\theta \; \d\phi^2\right).
\label{E:horizon}
\end{equation}
This is a restriction of cosmographic-LTB, (with $A(r,t)\to A(t)$). 
\end{itemize}
The running metrics (\ref{E:void}) and (\ref{E:wall}) are then matched to the radial sections
 of (\ref{E:horizon}) by regional conformal stretching factors,
 $\exp\left(2\Theta(t_\mathrm{void},r_\mathrm{void})\right)$ and $\exp\left(2\Theta(t_\mathrm{wall},r_\mathrm{wall})\right)$.
When bootstrapping timescape cosmographies to timescape cosmologies one does not apply the Einstein equations directly, but instead applies the Buchert averaging procedure~\cite{Wiltshire:2007-1, Wiltshire:2007-2, Wiltshire:2007-3, Leith:2007, Wiltshire:2008, Wiltshire:2009, Smale:2010, Wiltshire:2011-1, Smale:2011, Wiltshire:2011-2, Duley:2013, Wiltshire:2013}.

For all of these cosmographies one could introduce an extra, arbitrarily large, conformal factor $e^{2\Theta(t,\vec x)}$ without \emph{additionally} distorting the CMB,  beyond the \emph{intrinsic} distortional effects of these LTB, Swiss-cheese, and timescape cosmographies. 
In this regard, the CFLRW spacetimes are intrinsically simpler in that there simply are no low-redshift recent-universe CMB distortions to worry about, (at least until one adds the ``small perturbations''). This is why the CFLRW spacetimes are the main focus of this article.

%-------------------------------------------------------------------------------------------
\section{Redshift in CFLRW cosmography} 
%-------------------------------------------------------------------------------------------

Now in any CFLRW cosmography there will certainly be \emph{some} constraints on the conformal mode $\Theta(t,\vec x)$ arising from purely cosmographic observations. Consider for example the completely general formula for the (total) redshift~\cite{Stephani}:
\begin{equation}
1+z =
 {(g_{ab} \,k^a \,V^b)_\mathrm{emitter} \over (g_{ab} \,k^a \,V^b)_\mathrm{receiver} }.
\end{equation}
Here $V^a$ denotes the 4-velocity of the emitter/receiver, while the null vector $k^a$ is an \emph{affinely parameterized} tangent to the photon trajectory. A quite standard computation then leads to the factorized equation:
\begin{equation}
\label{E:redshift}
1+z =
{[\gamma \; (1 - \hat k \cdot \vec \beta)]_\mathrm{emitter}\over 
 [\gamma \; (1 - \hat k \cdot \vec \beta)]_\mathrm{receiver} } 
 \times { a_\mathrm{receiver} \over a_\mathrm{emitter} }
 \times  {e^{\Theta_\mathrm{receiver}}\over e^{\Theta_\mathrm{emitter}} }.
\end{equation}
Here the first two factors are completely standard, corresponding to the effects of peculiar motions of the emitter and receiver, and the standard cosmological redshift due to the overall expansion of the universe. The effect of peculiar motions can in the usual manner be further factorized and recast in terms of longitudinal and transverse contributions:
\begin{eqnarray}
\gamma \; (1 - \hat k \cdot \vec \beta) &=& {1-\beta_\parallel\over \sqrt{ 1-\beta_\parallel^2-\beta_\perp^2} }
= 
 \sqrt{ 1-\beta_\parallel^2\over 1-\beta_\parallel^2-\beta_\perp^2} \times 
\sqrt{1 - \beta_\parallel\over 1 + \beta_\parallel}.
\end{eqnarray}
Here the last factor is the usual longitudinal Doppler shift, and the first factor includes the effect of the transverse Doppler shift.

The ``new physics'' in the redshift equation~(\ref{E:redshift}) is  encoded in the third factor appearing therein. This extra factor, $\exp(\Theta_\mathrm{receiver} - \Theta_\mathrm{emitter})$, is due to the local conformal stretching. Its appearance can ultimately be traced back to the fact that  if $\d\lambda_0$ is an affine parameter for a null geodesic in the metric $g_0$, then  $\d \lambda = e^{2\Theta(t,\vec x)} \, \d\lambda_0$ is an affine parameter for the same null geodesic in the conformally related metric $g= e^{2\Theta(t,\vec x)}\; g_0$.
(In contrast, if  $\d\tau_0$ is the proper time for a  timelike curve in the metric $g_0$, then  $\d \tau = e^{\Theta(t,\vec x)} \, \d\tau_0$ is the proper time for the same timelike curve in the conformally related metric $g= e^{2\Theta(t,\vec x)}\; g_0$. It is this mismatch in exponents that leads to the final conformal term in the redshift formula.)

What purely observational theory-free constraints can we place on this new factor, $\exp(\Theta_\mathrm{receiver} - \Theta_\mathrm{emitter})$, which is now making an extra contribution to the redshift? First, note that because we typically do \emph{not} see wild variations in redshift \emph{within} individual galaxies, it must be that  $\exp(\Theta_\mathrm{emitter})$ is more or less constant \emph{within} individual galaxies, (more on this point later), though it can in principle vary significantly \emph{from galaxy to galaxy}.
If the conformal mode $\Theta$ does vary significantly from galaxy to galaxy then one would be mis-estimating galactic sizes and time-scales  by a factor $e^{\Theta_\mathrm{emitter}}$. While one cannot resolve individual solar systems within other galaxies, this could instead have some effect on estimates of internal galactic dynamics and the inferred need for dark matter --- see the discussion in section 7 below.

%-------------------------------------------------------------------------------------------
\section{Supernovae} 
%-------------------------------------------------------------------------------------------

Now consider the supernova data, but re-interpreted in a perhaps somewhat unusual manner: There is an outstanding un-modelled and physically ill-understood variance in the supernovae data of~\cite{SS:1998-1, SS:1998-2, SS:1998-3, Riess:2000, Tonry:2003, WoodVasey:2007, Davis:2007, Kowalski:2008, Rubin:2008}:
\begin{equation}
\Delta \mu \sim 0.13104 \;\; \hbox{stellar magnitudes}.
\end{equation}
This is normally attributed to ``intrinsic variability" of the supernovae, but let us see what happens if we instead attribute this (at least partially) to local variations in the conformal mode $\Theta_\mathrm{emitter}$.
Since
\begin{equation}
\mu = {5 \ln (d_L/ 1 \;\mathrm{Mpc})\over\ln 10}+ 25,
\end{equation}
it follows that any uncertainty in stellar magnitudes is related to an uncertainty in luminosity distance by
\begin{equation}
\Delta \mu = {5  \Delta [\ln (d_L/ 1\; \mathrm{Mpc})]\over\ln 10}.
\end{equation}
Thus
\begin{equation}
{d_L{}_\mathrm{,estimated} \over d_L{}_\mathrm{,true} } \lesssim \exp( \ln10 \times 0.13104/5) %= 10^{0.13104/5} 
= 1.06220,
\end{equation}
and
\begin{equation}
{d_L{}_\mathrm{,estimated} \over d_L{}_\mathrm{,true} } \gtrsim \exp(- \ln10 \times 0.13104/5) %= 10^{-0.13104/5} 
= 0.9414.
\end{equation}
We can re-phrase this as
\begin{equation}
{\Delta d_L \over d_L } \lesssim 6\%.
\end{equation}
Alternatively, in terms of absolute luminosity, we have
\begin{equation}
{\Delta L_* \over L_* } \lesssim 12\%.
\end{equation}
These are 1-sigma estimates. To now relate this to a possible local variability in the conformal factor $e^\Theta$ note that in a CFLRW universe, (ignoring peculiar motions, and after accounting for the overall expansion), the observed absolute luminosity is
\begin{equation}
[L_*]_\mathrm{observed} = e^{2\Theta_\mathrm{receiver}-2\Theta_\mathrm{emitter}}\; [L_*]_\mathrm{physical}.
\end{equation}
Here one factor of $e^{\Theta_\mathrm{receiver}-\Theta_\mathrm{emitter}}$ comes from the fact that the energy of each individual photon is redshifted, and another factor of  $e^{\Theta_\mathrm{receiver}-\Theta_\mathrm{emitter}}$ comes from the fact that physical clocks on the emitter are being slowed down with respect to standard FLRW clocks. Now, because $\Theta_\mathrm{receiver}$ is the same for each galaxy we look at,  $\Theta_\mathrm{receiver}$  drops out of the calculation,  from the SNAE--Ia data  we have
\begin{equation}
{\Delta [e^{\Theta_\mathrm{emitter}}] \over \langle e^{\Theta_\mathrm{emitter}}\rangle } \lesssim 6\%.
\end{equation}
This is a statement about the possible variations in $\Theta_\mathrm{emitter}$, not within individual galaxies, but over the ensemble of supernova-bearing regions of those various host galaxies that have survived all the data selection cuts. 
(This is a particularly tricky point --- any really ``stand out'' supernova with large $\Theta_\mathrm{emitter}$ would likely not have survived the data selection cuts. This is effectively a possible instance of confirmation bias;  any raw observational results that might be too far outside the expected range are likely to be culled as simple mistakes, and may not even be recorded. Any \emph{ex post facto} attempt at quantifying such possible data selection effects is almost certainly doomed; without direct access to the complete raw unprocessed data very little can be said.) 

%-------------------------------------------------------------------------------------------
\section{Dark energy} 
%-------------------------------------------------------------------------------------------

Without some independent argument we cannot yet say anything about the \emph{local}, (internal to a particular galaxy  or galactic cluster), averages $\langle e^{\Theta_\mathrm{emitter}}  \rangle$ or $\langle e^{\Theta_\mathrm{reciever}} \rangle$. (Though per definition we have either set the universe-wide spatial average to unity, $\langle e^{\Theta}  \rangle_\mathrm{spatial} = 1$, or made some equivalent gauge choice.) 

Some information in this regard can be extracted from estimates of the dark energy. One alternative to the standard picture of  cosmological dark energy is to suppose that the Hubble parameter we locally measure might for some reason be an over-estimate that is atypical of the universe as a whole~\cite{Wiltshire:2007-1, Wiltshire:2007-2, Wiltshire:2007-3, Leith:2007, Wiltshire:2008, Wiltshire:2009, Smale:2010, Wiltshire:2011-1, Smale:2011, Wiltshire:2011-2, Duley:2013, Wiltshire:2013, Wiltshire:2014}. But such behaviour is completely natural within the CFLRW models since, (with $\tau$ being the proper time of a specific observer), we have
\begin{equation}
H_\mathrm{local} = {\d a\over \d\tau } = {\d a\over \d t} {\d t \over \d\tau } 
= H_\mathrm{cosmological} \; e^{-\Theta_\mathrm{receiver}}. 
\label{E:hubble}
\end{equation}
So to eliminate dark energy we would need $\d t / \d\tau  > 1$, that is  $\d \tau  < \d t$, so that clocks need to be anomalously slow within galaxies (in particular, within our own galaxy the Milky Way). 
Specifically:
To completely eliminate the need for dark energy, the clocks in galaxy clusters, (and in particular in our own galaxy cluster), need to run about $\sqrt{3}$ times slower than clocks in voids.\footnote{Such an effect, arising from a conformal stretching of walls and filaments (containing galaxies) relative to voids, is also realized in the timescape scenario~\cite{Wiltshire:2007-1, Wiltshire:2007-2, Wiltshire:2007-3, Leith:2007, Wiltshire:2008, Wiltshire:2009, Smale:2010, Wiltshire:2011-1, Smale:2011, Wiltshire:2011-2, Duley:2013, Wiltshire:2013, Wiltshire:2014}. In that model the specific relation for the dressed and bare Hubble parameters differs from equation (\ref{E:hubble}), as it also involves a time derivative of $\exp{\Theta}$, see~\cite{Wiltshire:2007-2}. This allows phenomenologically realistic expansion histories to be obtained with local clock rate differences of (on average) 1.3 -- 1.4 at the present epoch~\cite{Leith:2007, Smale:2010, Duley:2013, Wiltshire:2014}, somewhat less than the bound of $\sqrt{3}$ estimated herein.}
(This estimate is based on the fact that standard cosmological models give a dark energy component with a mass that is about 3 times larger than that attributed to galaxy clusters, and noting that the mis-estimate of the cosmological energy density goes as the square of the mis-estimate of the Hubble parameter.) Note that the intra-galaxy variations of $\Theta$ could still be small; the $\sqrt{3}$ factor applies to the entire galaxy cluster.

%-------------------------------------------------------------------------------------------
\section{Dark matter} 
%-------------------------------------------------------------------------------------------

In contrast, attempting to eliminate the need for dark matter through this sort of mechanism is a considerably more radical proposal, and almost certainly non-viable. The point is that dark matter is associated with gravitationally bound galaxies, and to attribute all of the dark matter, (about 9 times the quantity of luminous matter), to a mis-calibration of clocks requires clocks in the spiral arms of galaxies to run some 3 times slower than in the outer reaches of the galactic halo. But one would then expect to see very large redshift variations across individual galaxies, contrary to observation. At best, one might hope to attribute some (relatively small) fraction of the dark matter to this mis-calibration effect. (There is again a particularly tricky issue with data selection cuts --- any really ``stand out'' galaxy with large internal variations of $\Theta_\mathrm{emitter}$ would likely not have survived usual data selection cuts.) 

So on the one hand these CFLRW cosmologies are quite radical proposals, while on the other hand we have just seen that, (even without appeal to dynamical considerations based on the Einstein equations), the situation is not completely unconstrained. However, it seems relatively difficult to develop additional direct (theory-independent) cosmographic tests that could rule such proposals out --- essentially because almost everything we know about the distant universe ultimately arises from our ability to count photons, and photon trajectories, (though not photon frequencies), are utterly insensitive to any otherwise unexpected conformal factor.

%-------------------------------------------------------------------------------------------
\section{CMB temperature} 
%-------------------------------------------------------------------------------------------
While the CFLRW cosmologies have been carefully constructed to avoid distorting the angular distribution, and angular temperature fluctuations, of the CMB photons, the locally measured net temperature is another matter. We have
\begin{equation}
T_\mathrm{CMB,local} = 
T_\mathrm{CMB,cosmological} \;\; e^{-\Theta_\mathrm{receiver}}. 
\end{equation}
This effect arises (from the redshift formula) because the wavelength of each CMB photon (equivalently its period)  is stretched by a factor $e^{\Theta_\mathrm{receiver}}$. (Recall that we had assumed $\Delta[\Theta_\mathrm{last~scattering}] \lesssim 10^{-5}$.)
The overall result of this effect is that we have now introduced the possibility that the CMB temperature (as measured by us) might differ significantly from the CMB temperature in voids. 
Furthermore, by adapting the supernova argument previously presented, the CMB temperature as measured by us might differ from the CMB temperature in the supernova-bearing regions of other galaxies by up to 6\%. 
(Again, modulo potential problems due to data selection cuts.)

%-------------------------------------------------------------------------------------------
\section{Dynamics} 
%-------------------------------------------------------------------------------------------

When attempting to move beyond cosmography, to develop an appropriate notion of CFLRW dynamics,  there is from a theoretical perspective almost an embarrassment of riches to draw on. 
A conformal mode in the spacetime geometry (effectively an extra scalar field) can be accommodated in any number of ways --- from various geometrically inspired ``conformal gravity'' models~\cite{Grumiller:2010, Maldacena:2011, Grumiller:2013}, through to various models that perform a ``change of frame'' to trade off the conformal mode  for some sort of scalar field~\cite{Capozziello:1996, Faraoni:1998} --- be it some form of inflaton, dilaton, axion, string modulus, or Higgs-like scalar field, through (more speculatively) to possibly invoking the difference between the ``string frame'' of particle physics from the ``Einstein frame'' of general relativity~\cite{Tong:2009}, through to various $f( R )$ modified theories of gravity~\cite{Sotiriou:2008, Aviles:2012, Aviles:2013, Capozziello:2014}, wherein a Legendre transformation can trade off the function $f(\_\!\_)$ for a self-interacting scalar field and its potential~\cite{Sotiriou:2008}. 
Even so, this is not necessarily an exhaustive list of available options for a dynamical analysis of the conformal mode~\cite{Bars:2013, Ohanian:2015}.
Unfortunately none of these approaches are quite as ``technically clean'' as one might wish for.

From a phenomenological perspective the modified gravity $f(R)$ theories~\cite{Sotiriou:2008, Aviles:2012, Aviles:2013, Capozziello:2014} are currently the most actively explored of these options. More generally, in any of these dynamical models the introduction of the extra degree of freedom in the conformal mode (scalar field) should be treated with extreme care --- there is always the potential of introducing a new fine-tuning problem. For instance, even if one were to kill off the need for dark energy (accelerated expansion) one would still need to explain why the observed/inferred quantum vacuum energy density is so small by particle physics standards: $\rho_\Lambda \lesssim 10^{-123} M_{Planck}/L_{Planck}^3$. 

Given the quite extensive range of possibilities listed above, I feel that focussing on any particular dynamical option would be at this stage premature. 
Instead it would seem advisable to spend some extra effort to develop additional  direct, (and as much as possible, theory-independent), cosmographic tests. Remember, (assuming the usual Einstein equations), that  97\% of the mass-energy of the universe is ``missing in action''.  Most of the current dynamical alternatives to dark energy/dark matter are rather ugly, so theory-free cosmographic tests hold considerable interest.

%-------------------------------------------------------------------------------------------
\section{Discussion} 
%-------------------------------------------------------------------------------------------
While traditionally the usual FLRW-based cosmological models have assumed that the cosmological spacetime geometry is smooth plus ``small perturbations'', I feel it is worthwhile to consider the possibility of large non-perturbative deviations from FLRW cosmology. Among all possible large non-perturbative deviations from FLRW cosmology, one class stands out as meriting particular attention: Conformal deformations of FLRW geometries are unique in that they permit significant inhomogeneity in the spacetime geometry \emph{without} damaging the smoothness of the cosmic microwave background --- at worst there will be an observer-dependent temperature shift. 
It is particularly important to carefully distinguish three different issues: The (known) smoothness of the CMB, the (hypothetical) smoothness of the geometry, and the (known lack of) smoothness of the matter distribution within the geometry. 

Supernova observations, and the concomitant luminosity distance estimates,  can then be used to constrain the variability of the conformal factor between the supernova-bearing regions of the relevant population of host galaxies in a manner that is largely independent of detailed cosmological assumptions. In contrast, theory-independent cosmographic constraints on the average value of the conformal factor are much weaker, and amount to the observation that the conformal factor cannot change appreciably over the visible disk of any individual galaxy. 

\enlargethispage{10pt}
In conclusion, in any conformally FLRW (CFLRW) cosmology one might quite reasonably hope to attribute the apparent existence of dark energy, (and possibly even some [presumably small] fraction of the dark matter), to a systematic mis-modelling of galactic and cosmological dynamics coming from an otherwise unexpected conformal factor. 
This general CFLRW framework thereby provides an extremely rich class of phenomenologically interesting models to consider, yet without permitting a dangerous uncontrolled free-for-all. It represents a minimalist non-perturbative extension of the standard FLRW cosmologies, a minimalist ``geometrically clumpy cosmology'', one that focusses attention on a reasonably specific set of rocks to look under.

%-------------------------------------------------------------------------------------------
\section*{Acknowledgments} 
%-------------------------------------------------------------------------------------------

This research was supported by a Marsden grant, and by a James Cook fellowship, both administered by the Royal Society of New Zealand. 

\noindent
I wish to thank John Barrow, Naresh Dadhich, Daniel Grumiller, and David Wiltshire for their comments, questions, and interest.

\clearpage
%-------------------------------------------------------------------------------------------
%-------------------------------------------------------------------------------------------
%-------------------------------------------------------------------------------------------

%-------------------------------------------------------------------------------------------
\end{document}